\documentclass[a4paper,11pt]{article}
\pdfoutput=1
\usepackage{pos}

\title{Electromagnetic form factors of the proton and neutron from \texorpdfstring{$N_f = 2 + 1$}{Nf=2+1} lattice QCD}
\author*[a]{Miguel Salg}
\author[b,c]{Dalibor Djukanovic}
\author[a]{Georg von Hippel}
\author[a,b,c]{Harvey B. Meyer}
\author[a]{Konstantin Ottnad}
\author[a,b,c]{Hartmut Wittig}

\affiliation[a]{\texorpdfstring{PRISMA${}^+$}{PRISMA+} Cluster of Excellence and Institute for Nuclear Physics, Johannes Gutenberg University Mainz, Johann-Joachim-Becher-Weg 45, 55128 Mainz, Germany}
\affiliation[b]{Helmholtz Institute Mainz, Staudingerweg 18, 55128 Mainz, Germany}
\affiliation[c]{GSI Helmholtzzentrum für Schwerionenforschung, 64291 Darmstadt, Germany}

\emailAdd{msalg@uni-mainz.de}

\abstract{
We present results for the electromagnetic form factors of the proton and neutron computed on the Coordinated Lattice Simulations (CLS) ensembles with $N_f = 2 + 1$ flavors of $\mathcal{O}(a)$-improved Wilson fermions and an $\mathcal{O}(a)$-improved conserved vector current.
In order to estimate the excited-state contamination, we employ several source-sink separations and apply the summation method.
The quark-disconnected diagrams entering the isoscalar quantities are computed explicitly.
For this purpose, a stochastic estimation based on the one-end trick is performed, in combination with a frequency-splitting technique and the hopping-parameter expansion.
By these means, we obtain a clear signal for the form factors including the quark-disconnected contributions, which have a statistically significant effect on our results.
From the $Q^2$-dependence of the form factors, we determine the electric and magnetic charge radii and the magnetic moments of the proton and neutron.
The chiral interpolation is carried out by simultaneously fitting the pion mass and $Q^2$-dependence of our form factor data directly to the expressions resulting from covariant chiral perturbation theory including vector mesons.
To assess the influence of systematic effects, we average over various cuts in the pion mass and the momentum transfer, as well as over different models for the lattice spacing and finite volume dependence.
\vspace{0.5cm}
\begin{flushright}
    MITP-22-076
\end{flushright}
}

\FullConference{%
The 39th International Symposium on Lattice Field Theory,\\
8th-13th August, 2022,\\
Rheinische Friedrich-Wilhelms-Universität Bonn, Bonn, Germany
}

\makeatletter
\hypersetup{pdftitle=\@title, pdfauthor=\printHeadAuthors}
\makeatother
\usepackage[USenglish]{babel}
\usepackage[group-digits=integer, exponent-product= \cdot]{siunitx}
\usepackage{csquotes}
\usepackage[english]{cleveref}
\usepackage[all]{foreign}
\usepackage{booktabs}
\usepackage{tikz}
\usepackage{braket}
\usepackage{subcaption}
\DeclareMathOperator{\tr}{tr}

\begin{document}
\maketitle

\section{Introduction}
The internal structure of the nucleon is still an open research field in subatomic physics.
In particular, there is a discrepancy between different measurements of the electric charge radius of the proton:
The value obtained from $ep$ scattering \cite{Bernauer2014}, while in good agreement with hydrogen spectroscopy \cite{Mohr2012}, is incompatible with the most accurate determination from the spectroscopy of muonic hydrogen \cite{Antognini2013}.
Hence, the electromagnetic form factors of the proton and neutron, from which the radius is extracted in the context of scattering experiments, are of lasting and high interest to the community.

For our theoretical calculations, we split the form factors into an isovector and an isoscalar part.
Whereas the former only contains quark-connected contributions, in the latter also quark-disconnected diagrams appear.
A full prediction of the proton and neutron form factors from first principles therefore necessitates a specific treatment of isoscalar quantities on the lattice, including the disconnected contributions.
Following our publication of the isovector electromagnetic form factors \cite{Djukanovic2021} and an initial study of the isoscalar ones \cite{Djukanovic2021a}, we present here the current state of our determination of the electromagnetic form factors of the proton and neutron from the $N_f = 2 + 1$ CLS ensembles \cite{Bruno2015}, where all contributions are evaluated explicitly.
Our preliminary results point towards a small value of the electric charge radius of the proton, consistent with the findings in Refs.\@ \cite{Djukanovic2021,Djukanovic2021a}.

These proceedings are organized as follows:
\Cref{sec:setup} describes our lattice setup and some computational details, while \cref{sec:analysis} is dedicated to the methods employed to extract the form factors and charge radii from our lattice data.
In \cref{sec:results} we present our preliminary results reflecting the current state of the analysis.
\Cref{sec:conclusions} draws some conclusions and gives an outlook to further planned work on this project.

\section{Lattice setup}
\label{sec:setup}
We use the CLS ensembles \cite{Bruno2015} which have been generated with $2 + 1$ flavors of non-perturbatively $\mathcal{O}(a)$-improved Wilson fermions \cite{Sheikholeslami1985,Bulava2013} and a tree-level improved Lüscher-Weisz gauge action \cite{Luescher1985}.
Only ensembles following the chiral trajectory characterized by $\tr M_q = 2m_l + m_s = \text{const.}$ are employed.
In order to prevent topological freezing, the fields obey open boundary conditions in time, with the exception of the ensembles E250 and D450, which use periodic boundary conditions in time.
\Cref{tab:ensembles} displays the set of ensembles entering the analysis:
They cover four lattice spacings in the range from \SI{0.050}{fm} to \SI{0.086}{fm}, and several different pion masses, including one slightly below the physical value (E250).
We note that data is available on more ensembles, but only those shown in \cref{tab:ensembles} are included in the final fits for this analysis.
Compared to last year's setup \cite{Djukanovic2021a}, we have added ensembles above but close to the physical pion mass, augmented the number of configurations and/or sources on several ensembles, and generated data at further source-sink separations on all of them.

\begin{figure}[ht]
    \begin{center}
        \begin{tikzpicture}[scale=0.57]
	\node[left] at (-2,0) {$x$};
	\node[right] at (2,0) {$y$};
	
	\filldraw (-2,0) circle (0.07) -- (2,0) circle (0.07);
	\draw (-2,0) to [out=45, in=180] (0,1) to [out=0, in=135] (2,0);
	\draw (-2,0) to [out=315, in=180] (0,-1) to [out=0, in=225] (2,0);
	
	\node[left] at (4.5,0) {$x$};
	\node[right] at (8.5,0) {$y$};
	\node[above] at (6.5,1) {$z$};
	
	\filldraw (4.5,0) circle (0.07) -- (8.5,0) circle (0.07);
	\draw (4.5,0) to [out=45, in=180] (6.5,1) to [out=0, in=135] (8.5,0);
	\filldraw[color=red] (6.5,1) circle (0.07);
	\draw (4.5,0) to [out=315, in=180] (6.5,-1) to [out=0, in=225] (8.5,0);
	
	\node[left] at (11,0) {$x$};
	\node[right] at (15,0) {$y$};
	\node[above] at (13,1.5) {$z$};
	
	\filldraw (11,0) circle (0.07) -- (15,0) circle (0.07);
	\draw (11,0) to [out=45, in=180] (13,1) to [out=0, in=135] (15,0);
	\draw (13,2) circle (0.5);
	\filldraw[color=red] (13,1.5) circle (0.07);
	\draw (11,0) to [out=315, in=180] (13,-1) to [out=0, in=225] (15,0);
	
	\node at (0,-2) {$C_2$};
	\node at (6.5,-2) {$C_3^\mathrm{conn}$};
	\node at (13,-2) {$C_3^\mathrm{disc}$};
\end{tikzpicture}
    \end{center}
    \caption{Diagrammatic representation of the two- and three-point functions of the nucleon. Only quark lines are shown, while all gluon lines are suppressed. The red dots in the three-point functions represent the operator insertion.}
    \label{fig:C2_3}
\end{figure}
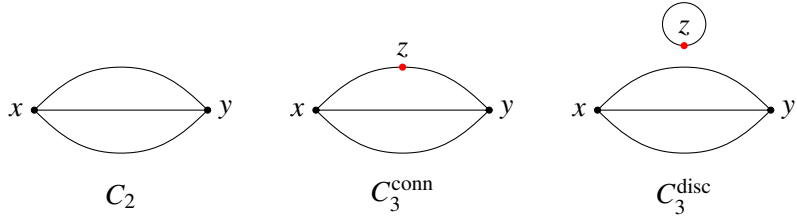

We measure the two- and three-point functions of the nucleon, which are depicted diagrammatically in \cref{fig:C2_3},
\begin{align}
    \label{eq:C2}
    \Braket{C_2(\mathbf{p'}; y_0, x_0)} &= \sum_{\mathbf{y}} e^{-i\mathbf{p'y}} \Gamma_{\beta\alpha} \Braket{0 | N_\alpha(\mathbf{y}, y_0) \bar{N}_\beta(\mathbf{0}, x_0) | 0} ,
\end{align}
\begin{align}
    \label{eq:C3_connected}
    \Braket{C_{3, O}(\mathbf{p'}, \mathbf{q}; y_0, z_0, x_0)} &= \sum_{\mathbf{y}, \mathbf{z}} e^{i \mathbf{qz}} e^{-i\mathbf{p'y}} \Gamma_{\beta\alpha} \Braket{0 | N_\alpha(\mathbf{y}, y_0) O(\mathbf{z}, z_0) \bar{N}_\beta(\mathbf{0}, x_0) | 0} , \\
    \label{eq:C3_disconnected}
    \Braket{C_{3, O}^\mathrm{disc}(\mathbf{p'}, \mathbf{q}; y_0, z_0, x_0)} &= \left\langle L^{O, \mathrm{disc}}(\mathbf{q}; z_0) C_2(\mathbf{p'}; y_0, x_0) \right\rangle , \\
    \label{eq:loops}
    L^{O, \mathrm{disc}}(\mathbf{q}; z_0) &= -\sum_{\mathbf{z}} e^{i \mathbf{q} \mathbf{z}} \tr[S(z, z) \Gamma] .
\end{align}

\begin{table}[ht]
    \centering
    \begin{tabular}{l@{\hspace{5mm}}cccccccc}
        \toprule
        ID   & $\beta$ & $a$ [fm]      & $N_\tau$ & $N_s$ & $M_\pi$ [MeV] & $N_\mathrm{cfg}^\mathrm{conn}$ & $N_\mathrm{cfg}^\mathrm{disc}$ \\ \midrule
        C101 & 3.40    & 0.08636(106)  & 96       & 48    & 224.9(3.0)    & 2000                           & 1000                           \\
        D450 & 3.46    & 0.07634(97)   & 128      & 64    & 216.3(2.8)    & 499                            & 499                            \\
        E250 & 3.55    & 0.06426(76)   & 192      & 96    & 129.1(1.7)    & 400                            & 400                            \\
        D200 & 3.55    & 0.06426(76)   & 128      & 64    & 203.0(2.5)    & 1999                           & 999                            \\
        E300 & 3.70    & 0.04981(57)   & 192      & 96    & 173.8(2.1)    & 569                            & 569                            \\
        J303 & 3.70    & 0.04981(57)   & 192      & 64    & 259.8(3.1)    & 1073                           & 1073                           \\ \bottomrule
    \end{tabular}
    \caption{Overview of the ensembles used in this study. The quoted errors on the pion masses include the error from the scale setting \cite{Bruno2017}.}
    \label{tab:ensembles}
\end{table}

Here, the same projection matrix $\Gamma = \frac{1}{2} (1 + \gamma_0) (1 + i \gamma_5 \gamma_3)$ is employed for both the two- and three-point functions, ensuring that the two of them are fully correlated.
In our setup, the nucleon at the sink is at rest, \ie for a momentum transfer \textbf{q} the initial and final states have momenta $\mathbf{p'} = \mathbf{0}$ and $\mathbf{p} = -\mathbf{q}$, respectively.
The disconnected part of the three-point functions is constructed from the quark loops and the two-point functions according to \cref{eq:C3_disconnected}.
The all-to-all propagator $S(z, z)$ appearing in the quark loops \cref{eq:loops} is computed via stochastic estimation using a frequency-splitting technique \cite{Giusti2019}.
To that end, we employ a hopping-parameter expansion for one heavy quark flavor and subsequently apply the one-end trick for the remaining flavors.
Furthermore, we average over the forward- and backward-propagating nucleon for the disconnected contribution.

To reduce the cost of the inversions, we apply the truncated-solver method \cite{Bali2010,Blum2013}.
In this context, additional measurements of the two-point function are used on the ensembles C101, D200, E300, and J303 to extend the statistics for the disconnected contribution.
For these additional measurements, we place the nucleon sources on different timeslices in the bulk of the lattice.
On all ensembles, we employ iterative statistics for the different source-sink separations.
This means that with rising $t_\mathrm{sep}$, the statistics for the connected part is increased.
For the disconnected part, the highest statistics at our disposal is always utilized, in order to get the best signal.

As in Ref.\@ \cite{Djukanovic2021}, we use a symmetrized conserved vector current, so that no renormalization is required.
The $\mathcal{O}(a)$-improvement is performed with the improvement coefficients computed in Ref.\@ \cite{Gerardin2019a}.
The remaining technical aspects of our setup are identical to our previous papers \cite{Harris2019,Djukanovic2021}, to which we refer the interested reader.

\section{Analysis procedure}
\label{sec:analysis}
To extract the form factors and charge radii from our lattice data, we proceed in three steps, which are presented in the following.

Starting from the two- and three-point functions \cref{eq:C2,eq:C3_connected,eq:C3_disconnected}, we calculate the ratios \cite{Korzec2009,Wilhelm2019}
\begin{align}
    R_O(\mathbf{p'}, \mathbf{q}; t_\mathrm{sep}, t) = \frac{\langle C_{3, O}(\mathbf{p'}, \mathbf{q}; t_\mathrm{sep}, t) \rangle}{\langle C_2(\mathbf{p'}; t_\mathrm{sep}) \rangle} \sqrt{\frac{\langle C_2(\mathbf{p'}-\mathbf{q}; t_\mathrm{sep} - t) \rangle \langle C_2(\mathbf{p'}; t) \rangle \langle C_2(\mathbf{p'}; t_\mathrm{sep}) \rangle}{\langle C_2(\mathbf{p'}; t_\mathrm{sep} - t) \rangle \langle C_2(\mathbf{p'}-\mathbf{q}; t) \rangle \langle C_2(\mathbf{p'}-\mathbf{q}; t_\mathrm{sep}) \rangle}} ,
    \label{eq:ratio}
\end{align}
where the source-sink separation is given by $t_\mathrm{sep} = y_0 - x_0$, and $t = z_0 - x_0$ denotes the temporal distance of the operator insertion from the source.
The two-point functions are averaged over equivalent momentum classes before plugging them into \cref{eq:ratio}.
At zero sink momentum, the effective form factors can be calculated from the ratios \cref{eq:ratio} by forming suitable linear combinations for different components of the vector current \cite{Wilhelm2019,Djukanovic2021}.

In general, baryonic correlation functions suffer from a strong signal-to-noise problem at large Euclidean time separations \cite{Lepage1989}.
This necessitates an explicit treatment of the excited-state systematics in order to extract the ground-state form factors from the effective ones computed at the typically accessible source-sink separations.
In this work, we employ the \enquote{plain} (one-state) summation method \cite{Capitani2015,Djukanovic2021}, where we vary the starting values $t_\mathrm{sep}^\mathrm{min}$ of the linear fits.
Rather than selecting one particular value of $t_\mathrm{sep}^\mathrm{min}$ on each ensemble as in Ref.\@ \cite{Djukanovic2021a}, we perform a weighted average over $t_\mathrm{sep}^\mathrm{min}$, where the weights are given by a smooth window function \cite{Djukanovic2022},
\begin{equation}
    \hat{G} = \frac{\sum_i w_i G_i}{\sum_i w_i} , \qquad w_i = \tanh\frac{t_i - t_w^\mathrm{low}}{\Delta t_w} - \tanh\frac{t_i - t_w^\mathrm{up}}{\Delta t_w} .
    \label{eq:window_average}
\end{equation}
Here, $t_i$ is the value of $t_\mathrm{sep}^\mathrm{min}$ in the $i$-th fit, and we choose $t_{w}^\mathrm{low} = \SI{0.8}{fm}$, $t_w^\mathrm{up} = \SI{1}{fm}$, and $\Delta t_w = \SI{0.08}{fm}$.
It should be stressed that the only quantity that is effectively restricted by this method is the minimal source-sink separation; all fits go up to the largest available $t_\mathrm{sep}$.

The charge radii are defined in terms of the $Q^2$-dependence of the form factors.
Instead of fitting each ensemble independently as in Ref.\@ \cite{Djukanovic2021a}, we adopt a procedure similar to the one presented in Ref.\@ \cite{Djukanovic2021}.
Namely, we combine the parametrization of the $Q^2$-dependence with the chiral and continuum extrapolation by performing a simultaneous fit of the pion mass and $Q^2$-dependence of our form factor data directly to the expressions resulting from covariant chiral perturbation theory \cite{Bauer2012}.
The fits are carried out for $G_E$ and $G_M$ simultaneously, but for the proton and neutron separately.
For both the proton and the neutron, we include the contributions from the $\rho$ meson in the expressions for the form factors.
For the neutron, we also include those from the $\omega$ and $\phi$ resonances, which improves the description of the data in this case.
The mass of the $\rho$ meson is set on each ensemble to the value at the corresponding pion mass and lattice spacing.
This is determined from a parametrization of the pion mass and lattice spacing dependence of a subset of the values for $M_\rho / M_\pi$ measured in Ref.\@ \cite{Ce2022}.
$G_E^p(0)$ is fixed by fitting the normalized ratio $G_E^p(Q^2) / G_E^p(0)$.
We perform several such fits with various cuts in the pion mass ($M_\pi \leq \SI{0.23}{GeV}$ and $M_\pi \leq \SI{0.27}{GeV}$) and the momentum transfer ($Q^2 \leq \SIrange[range-phrase = {, \ldots, }, range-units = single]{0.3}{0.6}{GeV^2}$), as well as with different models for the lattice spacing and/or finite volume dependence \cite{Djukanovic2021}, in order to estimate the corresponding systematic uncertainties.
If a parametrization of lattice artefacts is included, we stabilize the fit by means of Gaussian priors for the relevant coefficients.
For this purpose, we first perform fits to ensembles at $M_\pi \approx \SI{0.28}{GeV}$ only, where we have relatively precise data at a wide range of lattice spacings and volumes.
Here, we use a cut in $Q^2$ at \SI{0.6}{GeV^2} and a simultaneous description of the lattice spacing and finite volume dependence.
The coefficients for the correction terms as determined from these fits, together with their associated errors, are then employed as priors for the final fits to the ensembles listed in \cref{tab:ensembles}.

\section{Preliminary results}
\label{sec:results}
In the following, we present some illustrative results obtained so far with the procedures explained in \cref{sec:setup,sec:analysis}.

\begin{figure}[ht]
    \begin{center}
        \includegraphics[width=0.95\textwidth]{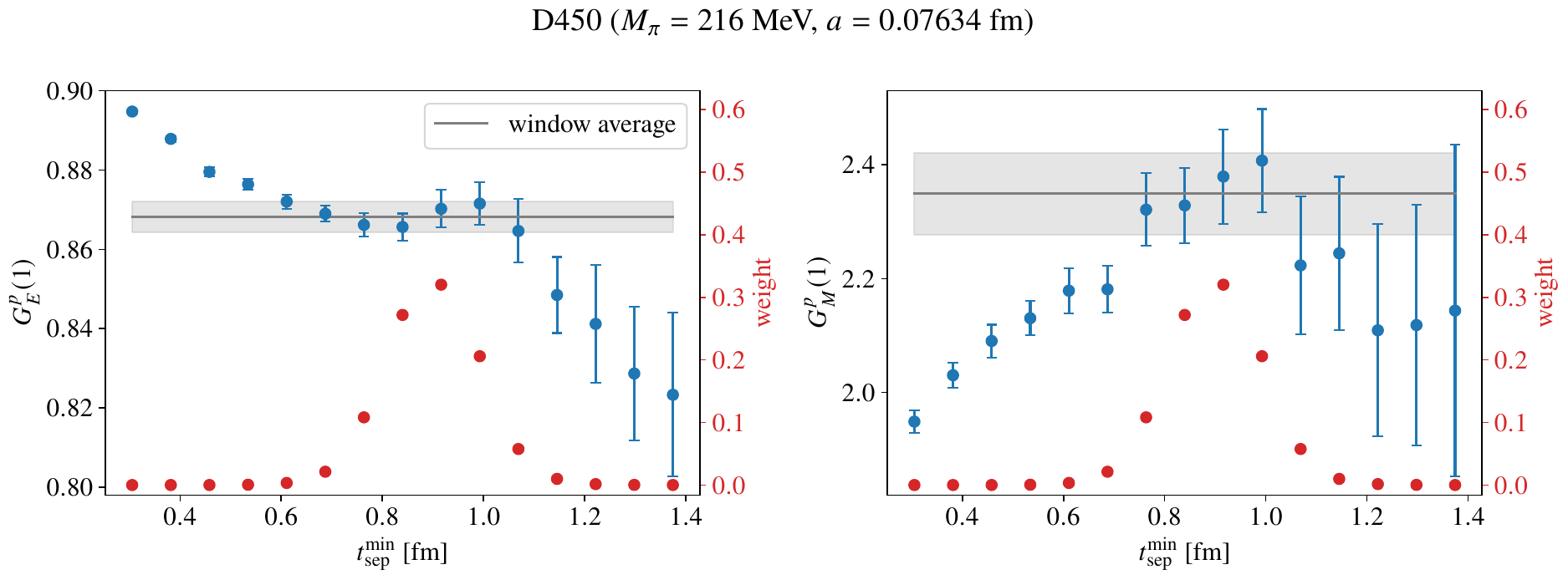}
    \end{center}
    \caption{Electromagnetic form factors of the proton at the first non-vanishing momentum on the ensemble D450 as a function of the minimal source-sink separation entering the summation fit. Each blue point corresponds to a single fit starting at the value given on the horizontal axis. The associated weights derived from \cref{eq:window_average} are represented by the red points, with the gray curves and bands depicting the averaged results.}
    \label{fig:window_average}
\end{figure}

For the effective form factors, we obtain a clear signal including the disconnected contributions.
Plotting them as a function of the operator insertion time, the curves for the different source-sink separations can be clearly distinguished in most cases.
The averaging of the results of the summation method over a smoothed window is shown in \cref{fig:window_average}.
One can see that the window average agrees well with the plateau visible by eye in the blue points.
This being valid on all ensembles, we conclude that the window method reliably detects the plateau with a reduced human bias, since we use the same window parameters in physical units on all ensembles.

\begin{figure}[ht]
    \begin{center}
        \begin{subfigure}{0.49\textwidth}
            \begin{center}
                \includegraphics[width=0.95\textwidth]{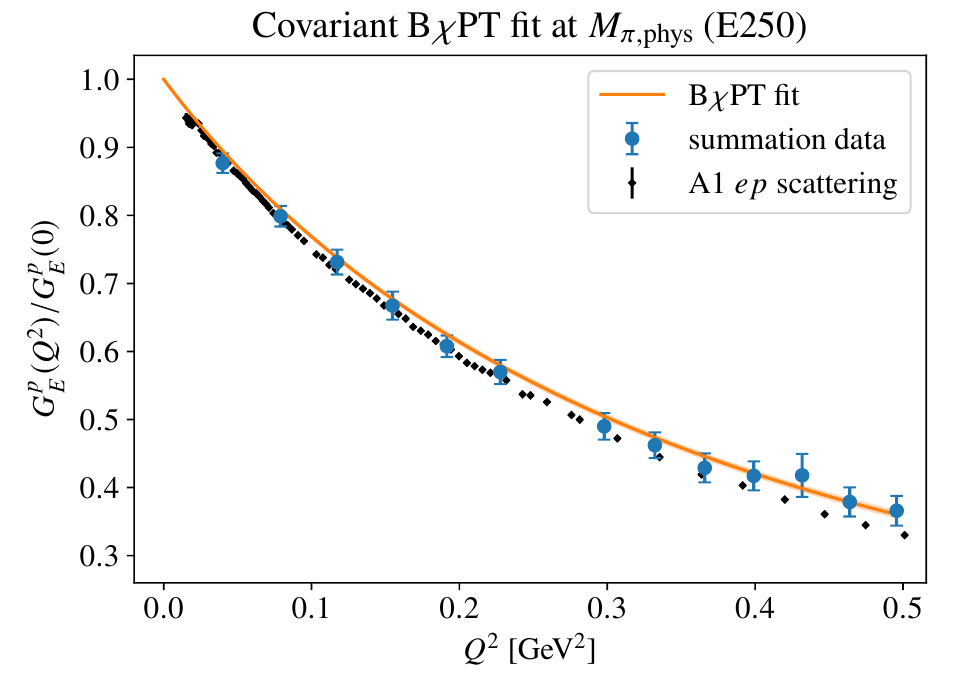}
            \end{center}
        \end{subfigure}
        \begin{subfigure}{0.49\textwidth}
            \begin{center}
                \includegraphics[width=0.95\textwidth]{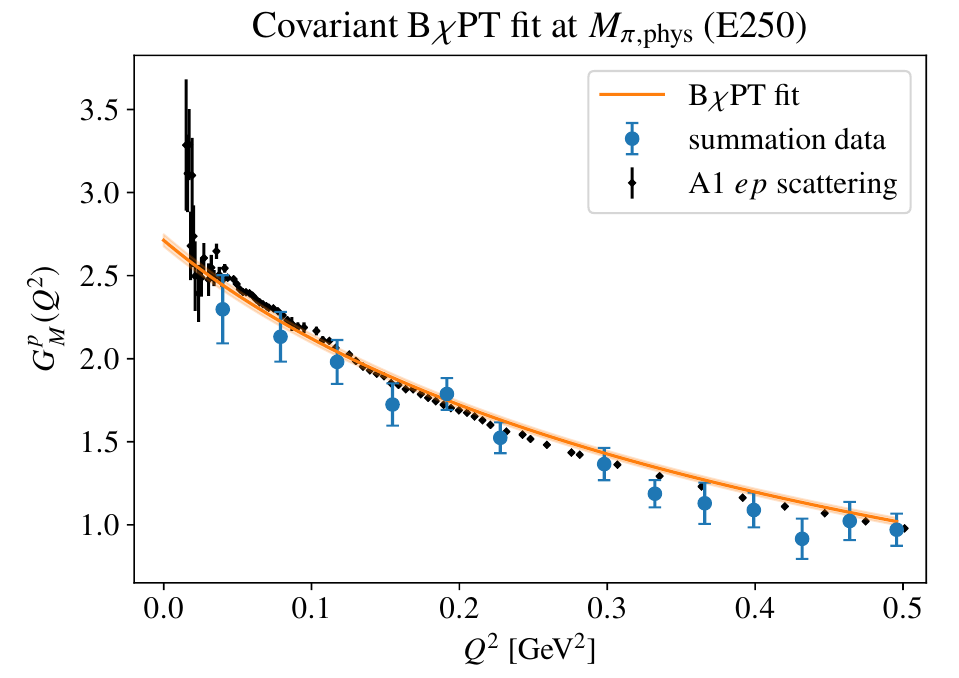}
            \end{center}
        \end{subfigure}
    \end{center}
    \caption{Electromagnetic form factors of the proton as a function of $Q^2$. Our lattice data as obtained from the summation method using the window average are represented by the blue points, with the orange curve and band depicting a B$\chi$PT fit with $M_{\pi, \mathrm{cut}} = \SI{0.23}{GeV}$ and $Q^2_\mathrm{cut} = \SI{0.5}{GeV^2}$. The black diamonds correspond to the experimental $ep$ scattering data from Ref.\@ \cite{Bernauer2014} obtained using Rosenbluth separation.}
    \label{fig:ge_gm_proton}
\end{figure}

\begin{figure}[ht]
    \begin{center}
        \begin{subfigure}{0.49\textwidth}
            \begin{center}
                \includegraphics[width=0.95\textwidth]{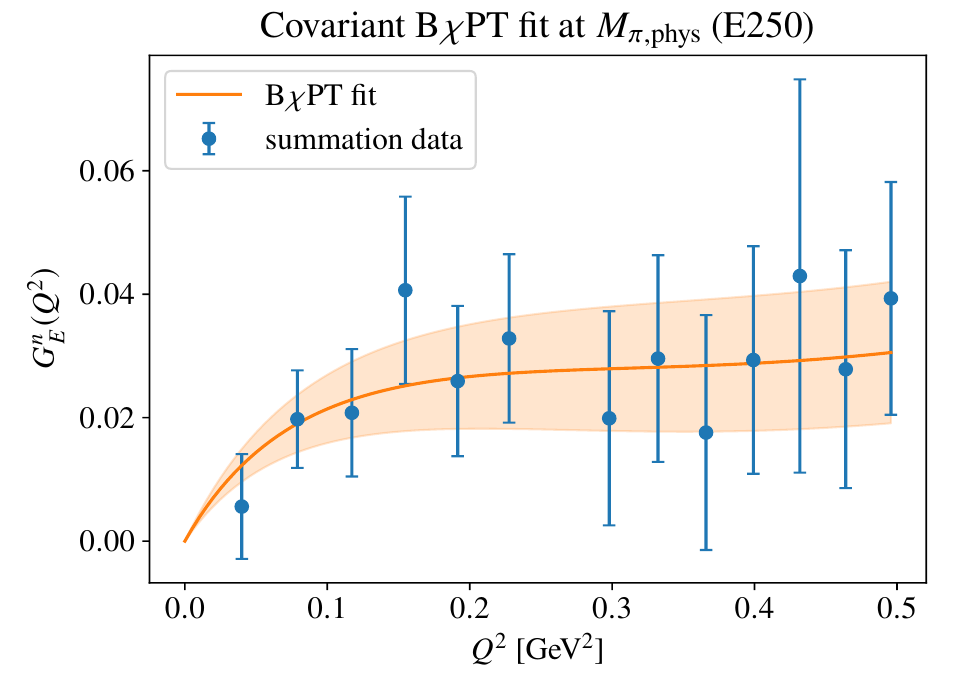}
            \end{center}
        \end{subfigure}
        \begin{subfigure}{0.49\textwidth}
            \begin{center}
                \includegraphics[width=0.95\textwidth]{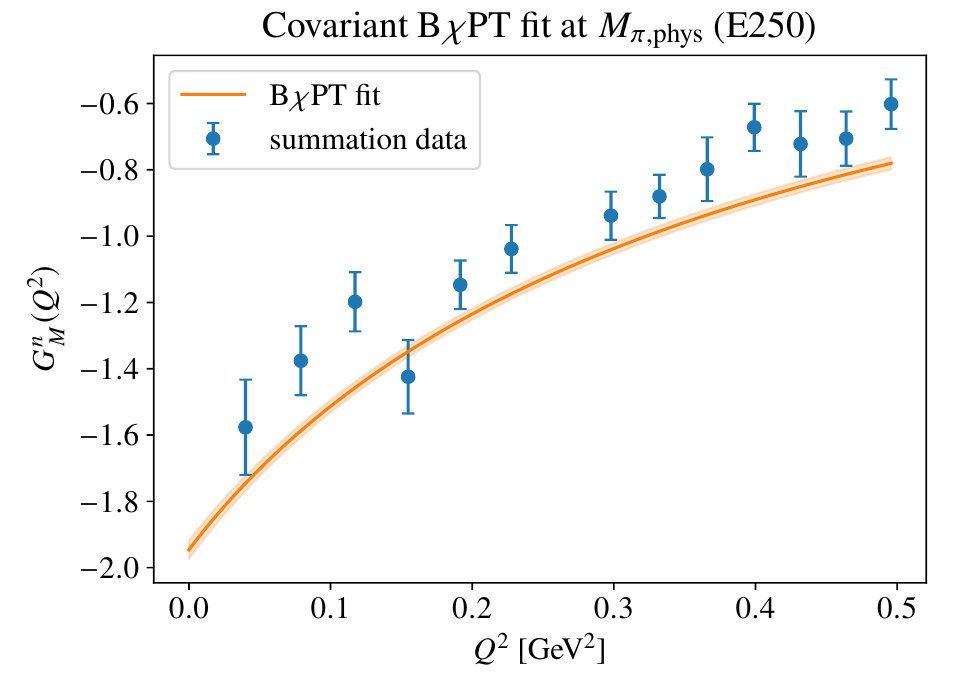}
            \end{center}
        \end{subfigure}
    \end{center}
    \caption{Electromagnetic form factors of the neutron as a function of $Q^2$. Our lattice data as obtained from the summation method using the window average are represented by the blue points, with the orange curve and band depicting a B$\chi$PT fit with $M_{\pi, \mathrm{cut}} = \SI{0.23}{GeV}$ and $Q^2_\mathrm{cut} = \SI{0.5}{GeV^2}$.}
    \label{fig:ge_gm_neutron}
\end{figure}

The $Q^2$-dependence of the form factors at the physical pion mass (E250) is displayed for the proton in \cref{fig:ge_gm_proton} and for the neutron in \cref{fig:ge_gm_neutron}.
These figures also feature a fit to baryon chiral perturbation theory (B$\chi$PT) yielding one of the best p-values.
In particular for the proton, the fit describes the data very well.
The drastically reduced error is due to the inclusion of several ensembles in one fit, with the data at larger pion masses being more precise than at $M_{\pi, \mathrm{phys}}$.
\Cref{fig:ge_gm_proton} includes furthermore the experimental data from $ep$ scattering \cite{Bernauer2014}, which agree with our results within our quoted errors in the region of small $Q^2$.
For the neutron, we find a slightly larger deviation between the fit and the data than for the proton, especially in the magnetic form factor.
Nevertheless, the p-value of the shown fit is acceptable, with $p \approx 0.12$.
For the electric form factor of the neutron, the relative uncertainties of the data are larger because of the absolute smallness of the quantity.
Still, the fit is able to describe the data reasonably well, with a correspondingly wider error band.

\begin{figure}[ht]
    \begin{center}
        \begin{subfigure}{0.49\textwidth}
            \begin{center}
                \includegraphics[width=0.79\textwidth]{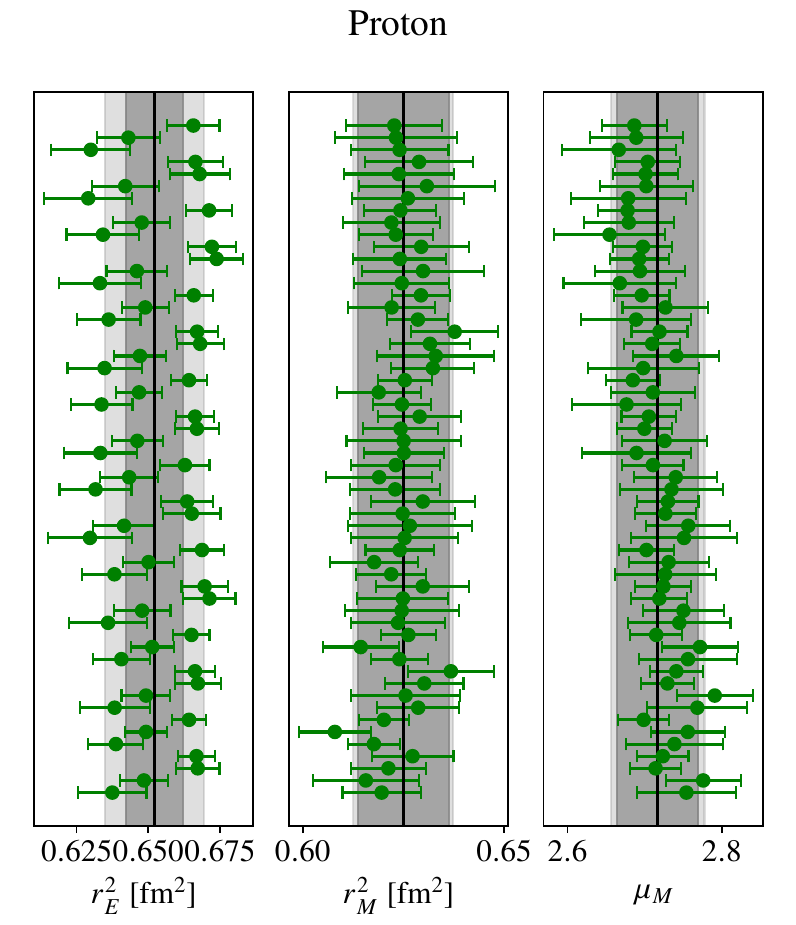}
            \end{center}
        \end{subfigure}
        \begin{subfigure}{0.49\textwidth}
            \begin{center}
                \includegraphics[width=0.79\textwidth]{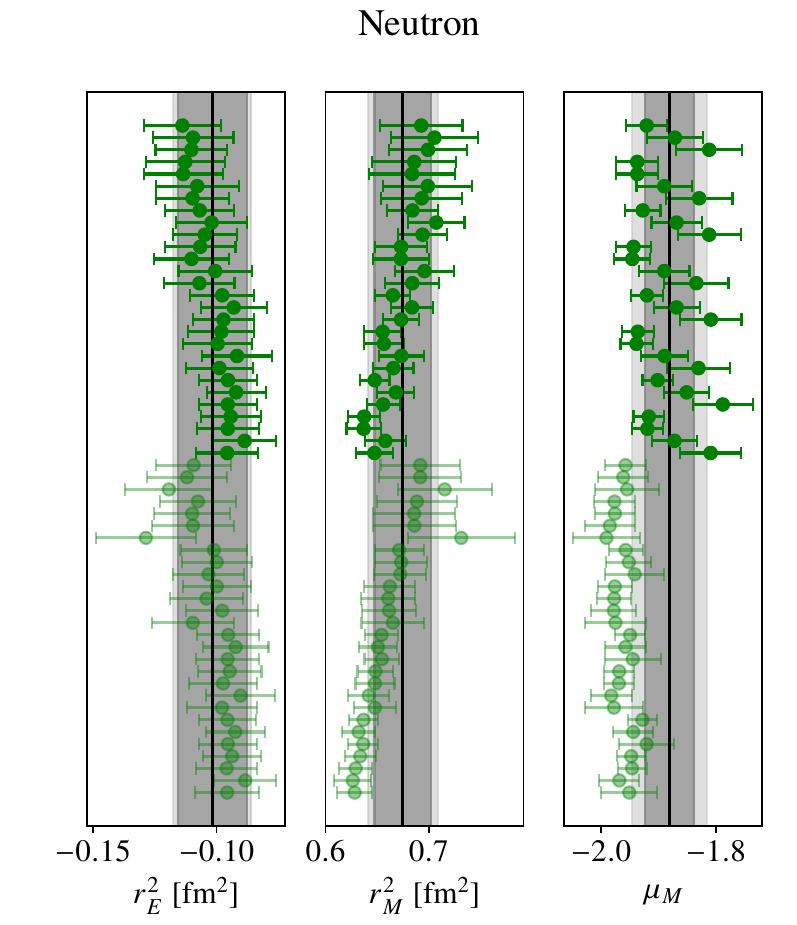}
            \end{center}
        \end{subfigure}
    \end{center}
    \caption{Electromagnetic charge radii and magnetic moments of the proton and neutron. The opaque green points depict the results of B$\chi$PT fits with a p-value of at least \SI{1}{\percent} (as determined from the augmented $\chi^2$), while the transparent ones originate from fits with a worse p-value and are excluded from the final average. The vertical black lines show the average central values and the inner (dark gray) bands the average statistical uncertainties. For the outer (light gray) bands, the systematic error estimates have been added in quadrature.}
    \label{fig:proton_neutron_radii_averaging}
\end{figure}

The collection of results for the electromagnetic charge radii and magnetic moments of the proton and neutron determined from these fits can be found in \cref{fig:proton_neutron_radii_averaging}.
For the proton, all fits have a p-value of at least \SI{1}{\percent}.
By contrast, the p-values for about half of the fits for the neutron are below that threshold.
For our final results, we perform naive (flat) averages after imposing a p-value cut at \SI{1}{\percent}.
We quote the average statistical uncertainty, and the variance determined from the spread of the fit results as a systematic error estimate \cite{Carrasco2014},
\begin{equation}
    \hat{x} = \frac{1}{N} \sum_{i = 1}^N x_i , \qquad \sigma^2_\mathrm{stat} = \frac{1}{N} \sum_{i = 1}^N \sigma^2_i , \qquad \sigma^2_\mathrm{syst} = \frac{1}{N} \sum_{i = 1}^N (x_i - \hat{x})^2 .
    \label{eq:naive_average}
\end{equation}

\begin{figure}[ht]
    \begin{center}
        \includegraphics[width=0.95\textwidth]{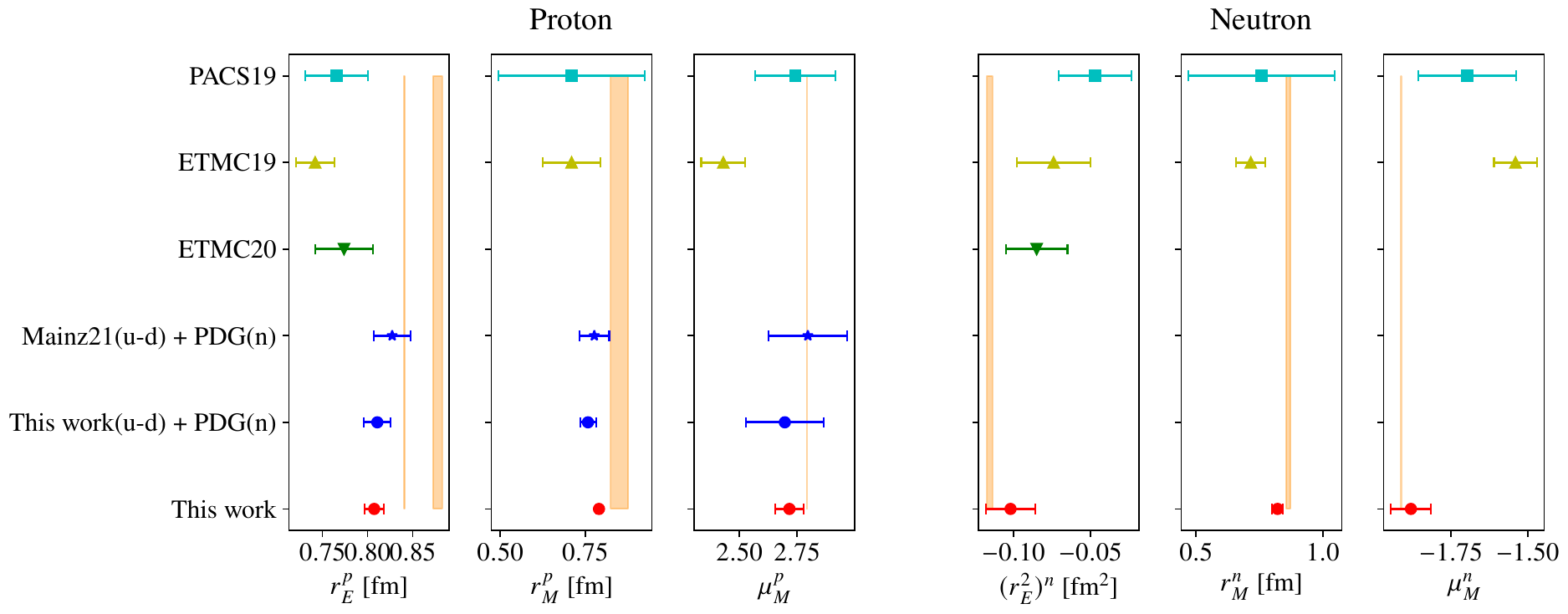}
    \end{center}
    \caption{Comparison of our preliminary results for the electromagnetic charge radii and the magnetic moments of the proton and neutron with other studies \cite{Djukanovic2021,Alexandrou2019,Alexandrou2020,Shintani2019}. The orange vertical bands depict the experimental values \cite{Mohr2012,Workman2022}.}
    \label{fig:proton_neutron_radii_comparison}
\end{figure}

In \cref{fig:proton_neutron_radii_comparison}, these results are compared to a selection of other studies:
Direct lattice determinations by PACS \cite{Shintani2019} and ETMC \cite{Alexandrou2019,Alexandrou2020}, the combination of our isovector data with the PDG values for the neutron (showing the results of our earlier published study, Ref.\@ \cite{Djukanovic2021}, or those from this analysis), and the experimental values \cite{Mohr2012,Workman2022}.
For the electric radius of the proton, we clearly favor a small value, as has been seen in previous lattice investigations.
For the other observables, our results lie within one to three standard deviations of the experimental findings, which is the kind of agreement one could expect from the other lattice determinations shown in \cref{fig:proton_neutron_radii_comparison}.
Furthermore, we achieve very competitive errors in particular for the radii, which is due to our direct fit strategy.

\section{Conclusions and outlook}
\label{sec:conclusions}
In these proceedings, we have investigated the electromagnetic form factors of the proton and neutron in lattice QCD with $2 + 1$ flavors of dynamical quarks including quark-connected and -disconnected contributions.
We have performed a chiral and continuum extrapolation by matching our lattice results with the predictions from covariant chiral perturbation theory.
From such combined fits of the pion mass and $Q^2$-dependence of the form factor data, the electromagnetic charge radii and magnetic moments of the proton and neutron have been extracted.
Our preliminary results agree well with the experimental values and previous lattice determinations, with very competitive errors, especially for the radii.
For the electric charge radius of the proton, they clearly point towards a small value.

In the future, it will be of great interest to study the influence of increased statistics for the disconnected contribution on our most chiral ensemble E250.
Besides, we are working on advanced averaging strategies for our fit results and the corresponding quantification of systematic uncertainties, as well as on a more complete understanding of the various trends seen in \cref{fig:proton_neutron_radii_averaging}.
Apart from that, some more details of the analysis procedure merit further investigation, in particular regarding the B$\chi$PT fits, which leave some room for potential improvement.

\acknowledgments
This research is partly supported by the Deutsche Forschungsgemeinschaft (DFG, German Research Foundation) through project HI 2048/1-2 and through the Cluster of Excellence \enquote{Precision Physics, Fundamental Interactions and Structure of Matter} (PRISMA${}^+$) funded by the DFG within the German Excellence Strategy.
Calculations for this project were partly performed on the HPC clusters \enquote{Clover} and \enquote{HIMster2} at the Helmholtz Institute Mainz, and \enquote{Mogon 2} at Johannes Gutenberg University Mainz (\url{https://hpc.uni-mainz.de}).
The authors also gratefully acknowledge the support of the John von Neumann Institute for Computing and Gauss Centre for Supercomputing e.V. (\url{https://www.gauss-centre.eu}) for projects CHMZ21, CHMZ36, NUCSTRUCLFL, and GCSNUCL2PT.

\newpage
\bibliography{literature.bib}
\end{document}